\documentclass[a4paper]{article}
\usepackage{amssymb}
\usepackage{graphicx}
\usepackage{color}
\usepackage[margin=1in]{geometry}

\newcommand{\RpPb}{\ensuremath{R_\mathrm{p-Pb}}}
\newcommand{\pt}{\ensuremath{p_\mathrm{T}}}

\title{Constraining nuclear Parton Density Functions with forward photon production at the LHC}
\author{Marco van Leeuwen\\
  \textit{Nikhef, National Institute for Subatomic Physics, P.O. Box 41882, 1009 DB Amsterdam}\\
   \textit{and Utrecht University, P.O. Box 80000, 3508 TA Utrecht,  The Netherlands} }

\begin{document}
\maketitle

\section{Introduction}
In this paper we explore the use of results on forward particle production at the LHC to constrain nuclear Parton Density Functions (nPDFs). The case study here is based on a possible future measurement of forward photon production with the  Forward Calorimeter that is currently under discussion as an upgrade of the ALICE experiment. As a starting point, we use the recent nNNPDF 1.0 nuclear PDFs, which have been determined using fixed target neutral-current DIS data, which constrain the gluon density at $x \gtrsim 10^{-2}$ \cite{AbdulKhalek:2019mzd}. The Bayesian reweighting technique is used to include the constraints from the future measurement.

\section{Method}
As a starting point, we use the recent nNNPDF 1.0 nuclear PDFs \cite{AbdulKhalek:2019mzd}, which have been determined using fixed target neutral-current DIS data, which constrain the gluon density at $x \gtrsim 10^{-2}$. The nNNPDF 1.0 nuclear PDFs provide a set of replicas which sample the allowed parton distributions in the nucleus. The reweighting process starts by calculating the photon production cross section in the rapidity range covered by the proposed measurement, $3.5 < \eta <4.5$ in the centre-of-mass system. The cross section is calculated using the INCNLO code \cite{Aurenche:1998gv}, with a patch to improve the numerical stability at forward rapidity, provided by Ilkka Helenius \cite{Helenius:2014qla} and interface code to read the PDFs using the LHAPDF library \cite{Buckley:2014ana}. The calculation is carried out using the nNNPDF 1.0 PDFs for the Pb nucleus and the NNPDF 3.1 \cite{Ball:2017nwa} PDF for the proton.

The reweighting itself follows the Bayesian approach outlined in \cite{Ball:2010gb,Gao:2017yyd}. For each replica $R_{i}$ the probability $P_i(\mathrm{data}|R_i)$ to observe the pseudo data points as provided by the ALICE-FOCAL collaboration is calculated and the reweighted (posterior) probability $P'_{i}$ for each replica is calculated using Bayes' theorem:
\begin{equation}
P'_i = 1/N_{rep} \frac{P_i(\mathrm{data}|R_i)}{\sum_i P_i(\mathrm{data}|R_i)}
\end{equation}
where the prefactor $1/N_{rep}$ is the (prior) probability for each replica before reweighting. The probabilities $P_i(\mathrm{data}|R_i)$ are given by the $\chi^2$ probability distribution
\[
P(\chi^2|N) = \frac{1}{2^{\frac{N}{2}}\Gamma\left(\frac{N}{2}\right)} \left(\chi^2\right)^{\frac{N}{2}-1} e^{-\frac{\chi^2}{2}}
\]
with as input the $\chi^2$ between the measured data points and the cross sections calculated with each replica $R_i$. The calculation of the $\chi^2$ which takes into account the correlated systematic uncertainties is discussed below around Eq. \ref{eq:chi2}.  

Confidence intervals for the final results are calculated by sorting the replicas (or cross sections) by value and summing the reweighted probabilities. For example, for the 90\% CL, the probabilities are summed until a cumulative probability of 0.05 is reached to find the lower bound of the confidence interval; further summing until 0.95 is reached determines the upper bound.

It has also been suggested to use an 'unweighting' procedure to generate a new set of equal-probability replicas \cite{Ball:2011gg}. This is useful if the results are to be distributed for further fits or theoretical predictions. However, for the present purpose, this step is not necessary.

\subsection{Treatment of the systematic uncertainties}

A key point in the calculation of the probabilities for the data is taking into account correlated systematic uncertainties. For simplicity, we have decided in this case to add the systematic and statistical uncertainties on the FOCAL pseudo data in quadrature for each data point, which amounts to assuming that the systematic uncertainties are uncorrelated from point to point, and to add a fully correlated uncertainty of $\sigma_N = 5\%$, to represent normalisation uncertainties from the total cross section determination in the experiment, the $T_{pA}$ scaling factor for nuclei, and the energy scale uncertainty from the experiment. The effect of the correlated uncertainty is taken into account by introducing a parameter $\epsilon$ which quantifies the systematic deviation. The $\chi^2$ deviation of the data from the curves is then calculated as follows:
\begin{equation}
  \chi^2(\epsilon) = \sum_i {\left(\frac{(1-\epsilon \sigma_N)y_{i} - f(x_i)}{\tilde{\sigma}_{i}}\right)^2 } + \epsilon^2
  \label{eq:chi2}
\end{equation}
where the measured data points $(x_i,y_i)$ are shifted by a relative amount $\epsilon \sigma_{N}$. The function $f(x_i)$ is the differential cross section calculated for \pt{} bin $i$ of the pseudo data. The uncertainties on the data points $\sigma_i$ are rescaled, assuming that the relative uncertainty is constant $\tilde{\sigma}_i = (1-\epsilon \sigma_{N}) \sigma_i$.

A minimization algorithm is used to find the value of $\epsilon$ for which the $\chi^2$ is the smallest. This value of the $\chi^2$ is then used to calculate the unnormalized posterior probability $P(\mathrm{data}|R_j)$ from the  $\chi^2$ probability distribution with number of degrees of freedom equal to the number of data points minus one.

\section{Results}

\begin{figure}
  \centering
  \includegraphics[width=0.6\textwidth]{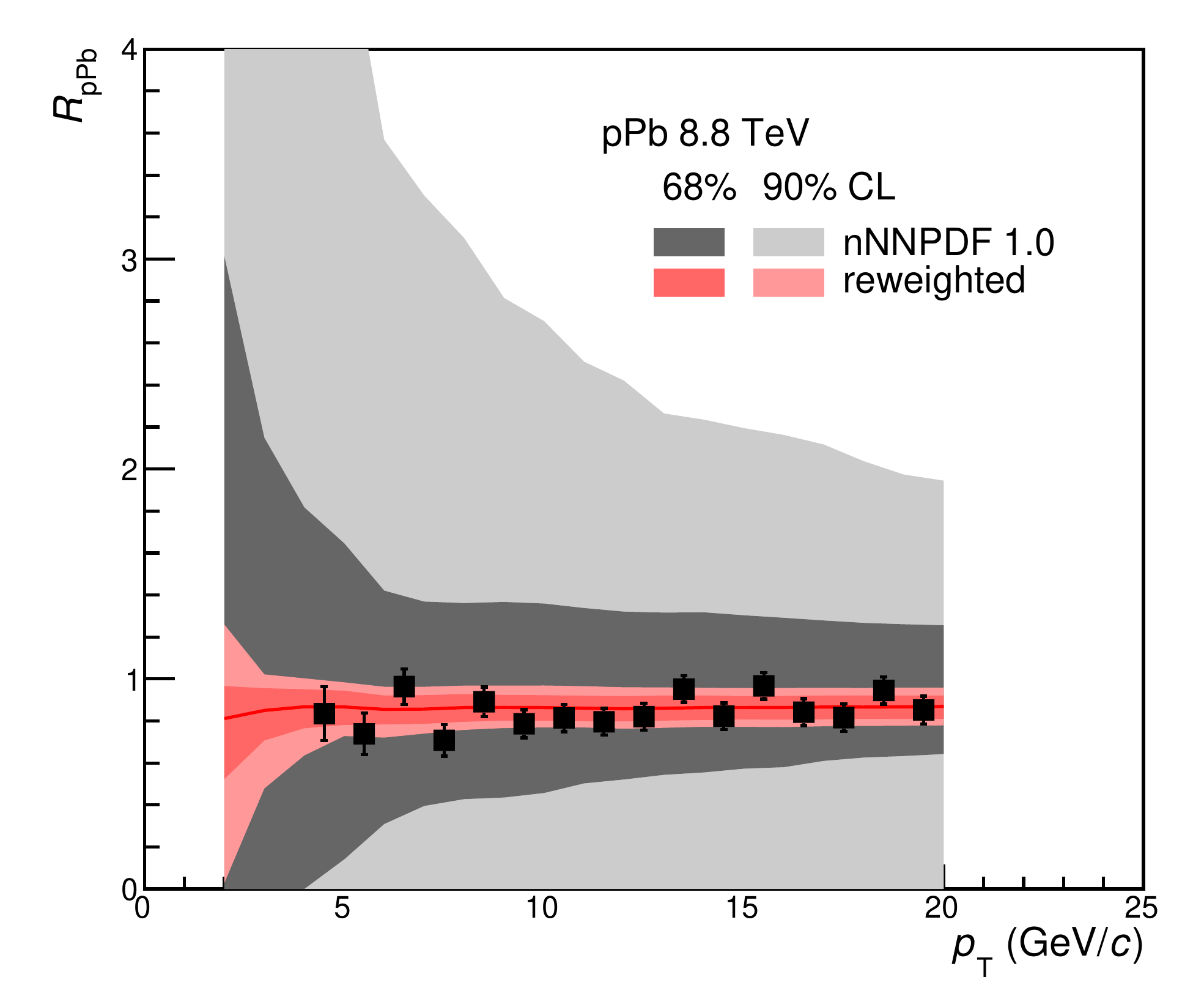}
\caption{\label{fig:rppb}Nuclear modification factor \RpPb{} for direct photons at $\eta=4.0$. The grey bands indicate the 68 and 90\% CL intervals from the nNNPDF 1.0. The points are pseudo data reflecting the projected uncertainties on the direct photon measurement using FoCal and the colored bands show the 68 and 90 \% CL intervals of the \RpPb{} calculations that are compatible with the pseudo data.}
\end{figure}

Figure \ref{fig:rppb} shows the nuclear modification factor of photons at forward rapidity $3.5 < \eta < 4.5$ calculated using the current NNPDF 1.0 and NNPDF 3.1 for the proton PDFs. The figure shows the theoretical calculations, with 68\% and 90\% CL intervals, the pseudo data from the ALICE-FOCAL collaboration and the posterior uncertainties, also depicted as a 68\% and 90\% CL intervals (in red). The uncertainties on the NNPDF are large, because this observable probes $x < 10^{-4}$ and the DIS measurements that constrain the NNPDFs are in the range $x > 10^{-2}$. Note also that he probability distribution is very non-Gaussian, as can be seen from comparing the 68\% and 90\% CL intervals. A measurement of forward photon production with the precision foreseen for the FoCal clearly provides experimental input to constrain the nuclear PDFs.

\begin{figure}
  \centering
  \includegraphics[width=0.6\textwidth]{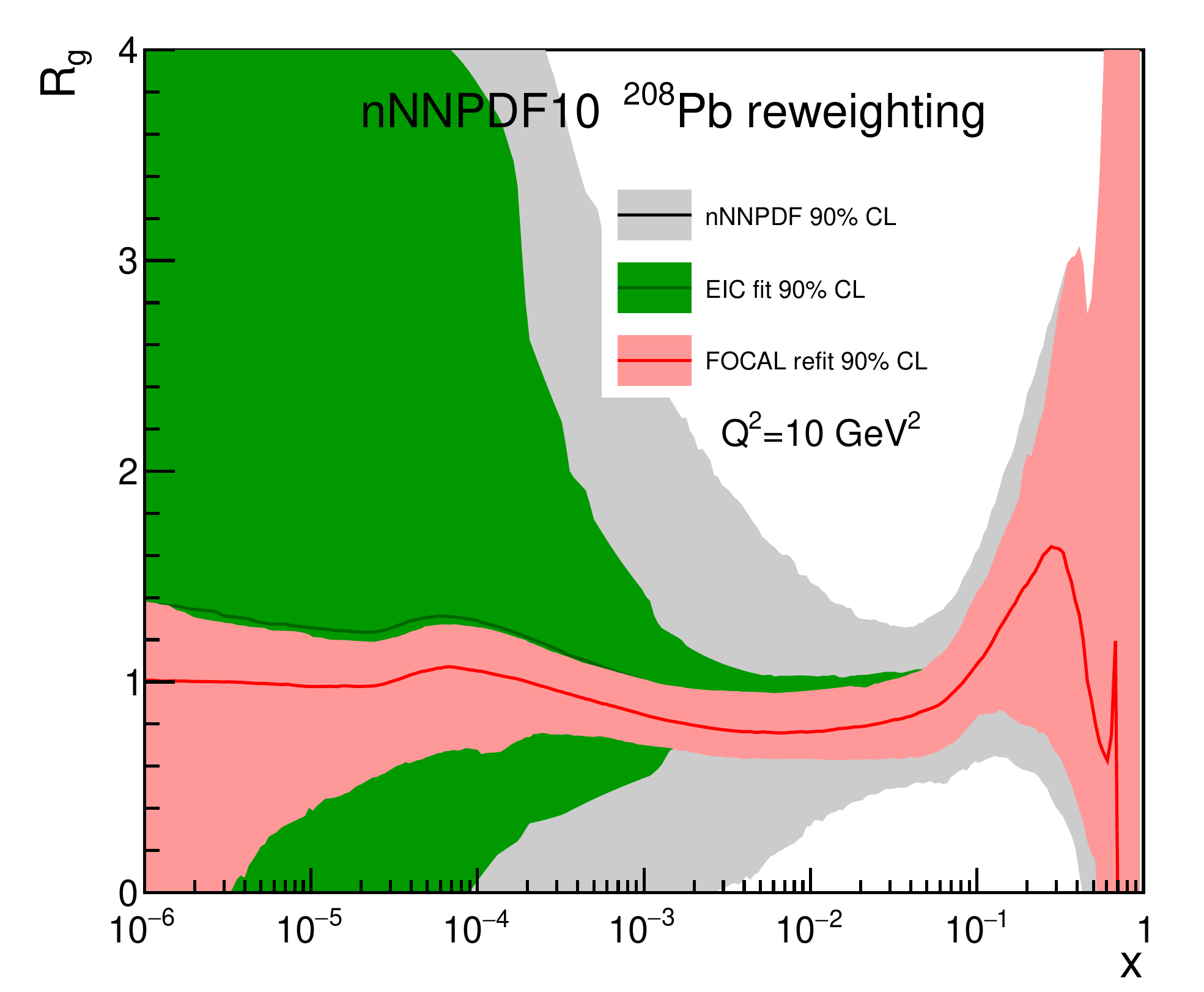}
\caption{\label{fig:Rg}Nuclear modification factor $R_{g}(x)$ for the gluon distribution in the Pb nucleus at a scale $Q^{2} = 10\;\mathrm{GeV}^2$. The grey bands indicate the 68 and 90 \% CL intervals from nNNPDF 1.0. The red colored bands show the 68 and 90 \% CL intervals of the posterior PDFs constrained by the foreseen FoCal measurement.}
\end{figure}

Figure \ref{fig:Rg} shows the resulting posterior distributions for the nuclear modification of the gluon PDF at a scale $Q^{2} = 10 \mathrm{GeV}^2$. For comparison, also the constraints provided by possible future measurements at the high energy EIC, as presented in the nNNPDF paper \cite{AbdulKhalek:2019mzd}, are shown. It can be seen that the forward photon measurement at the LHC constrains the gluon nPDF down to $x \approx 5 \cdot 10^{-5}$.

In addition to direct photon measurements, other forward particle production measurements are also sensitive to the gluon density at small-$x$. In particular, forward open charm measurements are already available and are being used to constrain both proton and nuclear PDFs \cite{Gauld:2016kpd,Eskola:2019bgf}. The different processes cover different kinematic ranges and have different theoretical uncertainties, and thus provide complementary tests of the small-$x$ structure of the nucleus.

\section*{Acknowledgments}
The author would like to thank the nNNPDF team (Jake Ethier, Juan Rojo and Rabah Abdul Khalek) for useful discussion, including help with the reweighting procedure, and for providing the 1000-replica version of the nNNPDF1.0. Thanks also to  ALICE-FoCal collaboration for providing the pseudo data.

\bibliography{nnpdf_photons}{}
\bibliographystyle{plain}
\end{document}